\documentclass[11pt]{article}
\usepackage{amssymb}

\usepackage{amsmath}
\usepackage{graphicx}

\renewcommand{\vec}[1]{{\bf #1}}       
\def\beq{\begin{eqnarray}}    
\def\eeq{\end{eqnarray}}      

\renewcommand{\Re}{\,\mbox{Re}\,}       

\def\al{\alpha}
\def\be{\beta}

\def\ga{\gamma}
\def\de{\delta}

\def\ze{\zeta}

\def\La{\Lambda}
\def\la{\lambda}
\def\na{\nabla}
\def\pa{\partial}

\def\si{\sigma}

\def\ph{\varphi}

\def\Ga{\Gamma}
\def\De{\Delta}
\def\La{\Lambda}

\begin{document}

\hfill DF/UFJF-01/02

\hfill August, 2002

\vskip 16mm

\begin{center}

{\Large\sc 
On the stability of the anomaly-induced inflation}
\vskip 12mm

{\bf A.M. Pelinson $^{a,b}$} 
 \footnote{Electronic address: ana@fisica.ufjf.br},
$\,$ $\,$
$\,$ $\,$
{\bf I.L. Shapiro $^{b,c}$}
 \footnote{Electronic address: shapiro@fisica.ufjf.br}
$\,$ $\,$
$\,$ $\,$
{\bf F.I. Takakura $^{b}$}
 \footnote{Electronic address: takakura@fisica.ufjf.br}
\vskip 12mm

a. {\small\sl 
Departamento de Campos e Part\'{\i}culas, CBPF, Rio de Janeiro, Brazil}
\vskip 4mm

b. {\small\sl Departamento de F\'{\i}sica -- ICE, 
Universidade Federal de Juiz de Fora, MG, Brazil}
\vskip 4mm

c. {\small\sl Tomsk State Pedagogical University, Tomsk, Russia}
\end{center}
\vskip 8mm
\vskip 8mm

\begin{quotation}
\noindent
{\large\it Abstract}.$\,\,$
We analyze various phases of inflation based on the 
anomaly-induced effective action of gravity (modified 
Starobinsky model), taking the cosmological constant $\Lambda$ 
and $k=0,\pm 1$ topologies into account. The total number of 
the inflationary $e$-folds may be enormous, but at the last 
65 of them the inflation greatly slows down due to the 
contributions of the massive particles. For the supersymmetric 
particle content, the stability of inflation holds from the 
initial point at the sub-Planck scale until the supersymmetry 
breaks down. After that the universe enters into the unstable 
regime with the eventual transition into the stable FRW-like 
evolution with small positive cosmological constant. It is 
remarkable, that all this follows automatically, without 
fine-tuning of any sort, independent on the values of 
$\Lambda$ and $k$. Finally, we consider the stability 
under the metric perturbations during the last 65 $e$-folds 
of inflation and find that the amplitude of the ones with the 
wavenumber below a certain cutoff has an acceptable range.
\end{quotation}
\newpage


\section{Introduction.}

$\,\,\,\,\,\,\,$
The importance of inflation becomes more and more apparent in 
the last years (see, e.g. \cite{Guth} for the review). One of 
the reasons is that the existing and future astrophysical data 
may give a chance to use the early universe as a laboratory 
for the high energy physics, providing the Theoretical 
Physics a new important information. However, in 
order to use this opportunity, one needs a model of inflation 
which should be free of any serious phenomenological input.
Otherwise, the experimental data will mainly deal with this 
input and not with the underlying high energy physics. 

In this paper, we are going to discuss the example of such a 
model - the new version \cite{anju,insusy,shocom} of the 
anomaly-induced inflation (Starobinsky model) 
\cite{fhh,star,star1,much,vile,ander}. This model 
is based on the well known methods of quantum field 
theory in curved space-time (see, e.g. \cite{birdav,book}) 
and on the effective quantum field theory approach. The 
advantage of the anomaly-induced inflation is that it 
does not contain usual elements of the conventional cosmological 
phenomenology such as inflaton, the precise form of its 
potential and/or the necessity of a fine-tuning for the 
initial data. The anomaly-induced inflation provides the
explanation for both the start of inflation and the graceful 
exit to the post-inflationary FRW-like phase \cite{insusy}. 
Under a natural supposition about the supersymmetry breaking 
scale we arrive at the magnitude of the Hubble constant which 
does not generate unacceptable amplitudes of the gravitational 
waves \cite{shocom}. 

The problems of stability play a great role in the 
anomaly-induced inflation. In particular, 
the transition between the stable and unstable regimes,
which could take place due to the breaking of supersymmetry 
\cite{insusy} and the quantum vacuum effects of massive fields
\cite{shocom}, could be responsible for the graceful exit 
to the FRW evolution. In the next sections, we investigate 
the stability with respect to the small perturbations of 
the conformal factor for various regimes associated to the 
anomaly-induced inflation. One can distinguish three stages
of the expansion:

$i)$ For the almost exponential inflation that takes place 
at the very beginning (when the effect of particle
masses is negligible) we consider the spaces with zero, 
positive and negative space curvatures, and with 
zero, positive or negative cosmological constant. The
main part of the results can be obtained analytically  
such that some very important general features are 
established.

$ii)$ At the post-inflationary stage, after the 
supersymmetry breaking, the universe is approaching the stable 
fixed point with a small Hubble constant. In this case the FRW 
solution is a very good approximation for the theory with 
vacuum quantum corrections, such that these 
corrections do not play an essential role at the later time. 
We shall verify that the FRW-like solution is an attractor 
for the theory with quantum corrections if the cosmological 
constant has positive sign.

$iii)$  The most complicated is the intermediate transitional 
epoch between the inflationary and the post-inflationary 
evolutions. This period is characterized by the vacuum 
quantum effects of both massless and massive fields. 
Indeed, the exact analytic 
methods for the stability analysis \cite{star} do not apply 
in this case, and one has to use numerical and approximate 
analytical methods.

The paper is organized as follows. In the next section, we 
present basic concepts of the anomaly-induced inflation. In 
part, we go beyond the results of the previous publications 
\cite{anju,insusy,shocom}. In particular, we take into account
the effect of the renormalization of the cosmological constant 
(here, our approach is somehow similar to the one of \cite{nova}). 
In section 3 the stability under the small perturbations of 
the conformal factor is analyzed for the ``massless'' stage $i)$, 
mainly through the analytical methods. The main new aspect 
here is that we clarify the role of the cosmological constant. 
In section 4 we verify, using both approximate analytical 
and numerical methods, that the stability of the model with 
the supersymmetric particle content really holds until the 
scale of the supersymmetry breaking. In section 5 we discuss, 
using the approximate analytical methods tested in section 4 
and also the numerical integrations, the stability under the
metric perturbations. Finally, in section 6 we draw our 
conclusions and also present a general discussion about the 
possible theoretical significance of the model.

\section{Brief survey of the anomaly-induced inflation}

$\,\,\,\,\,\,\,$
Consider the vacuum quantum effects in the Early Universe,
below the Planck scale. Then, the appropriate framework 
is not a string theory but some effective quantum field 
theory on a classical curved background. Let us make some 
assumptions about a quantum field theory which is valid at 
the sub-Planck energies. It proves useful to concentrate 
on the asymptotically free theories,  
that helps to avoid the discussion of the higher-loop effects. 
Thus, in what follows we shall deal with the one-loop 
approximation and only mention possible effects of the
higher loops. 
 
Consider the theory which has
$N_0$ real scalars, $N_{1/2}$ Dirac spinors, and $N_1$ vectors. 
One has to notice that the vacuum quantum effects originate 
from the virtual particles. Therefore, the numbers 
$\,N_0,\,N_{1/2},\,N_1\,$ correspond to the particle 
content of the quantum theory, but they do not describe 
the real matter which might fill the Universe. As we shall 
see later on, the real matter has little importance for the 
anomaly-induced inflation. For simplicity, one can assume
that the initial universe is empty and that all matter content
of the Universe is created during the reheating period 
\cite{vile}.

The renormalizable quantum theory of matter fields on 
curved background includes the action of vacuum, and
its minimal necessary form is
\beq
S_{vacuum}\, =\, S_{HD}\, + \,S_{EH}\,.
\label{vacuum}
\eeq
Here the first term contains higher derivatives of the metric
\beq
S_{HD}\, =\, \int d^4x\sqrt{-g}\,
\left\{ a_1 C^2 + a_2 E + a_3 {\nabla^2} R + a_4 R^2 \right\}
\label{higher}
\eeq
and the second is the Einstein-Hilbert term
\footnote{Our notations are $\eta_{\mu\nu}=diag(+---)$ and 
$\,\,R_{\mu\nu}=\partial_\lambda\,\Ga^\la_{\mu\nu}
-\,...\,\,.\,\,$ The sign 
of the cosmological term in (\ref{Einstein}) corresponds to
the positive vacuum energy for positive $\La$.}
\beq
S_{EH}\, =\,
-\,\frac{1}{16\pi G}\,\int d^4x\sqrt{-g}\,(R + 2\La)\,,
\label{Einstein}
\eeq
where
$\,a_1,..,a_4$, $\,G\,$ and $\,\La\,$ are the parameters of the 
vacuum action. The renormalization of these parameters and 
the quantum corrections to the action (\ref{vacuum}) are the 
basics of the inflationary model we are discussing here.
The discussion in this section will perform in two stages: 
first we review the massless limit and then present a useful 
approximation \cite{shocom} for the massive case.


\subsection{Massless conformal case}

$\,\,\,\,\,\,\,$
At high energies the masses of the fields are negligible
and massless conformal fields serve as a good approximation.
The vacuum quantum effects correspond to the Feynman diagrams
with a loop of the quantum matter field and external
lines of the classical gravity. Then, in order to evaluate 
the relevance of the mass of the quantum field, one has to 
compare it to the energy corresponding to the external line. 
In this paper we shall follow Ref. \cite{nova} and associate 
the energy of a quanta of the gravitational field with the 
magnitude of the Hubble constant $\,H$. Therefore, the massless 
approximation is justified for the very large values of $\,H$. 
As we shall see in the next subsection, 
this corresponds to the initial stage of inflation. 

The actions of the free real scalar,
Dirac fermion and gauge vector fields are 
\beq
S_s\,=\,\frac12\,\int d^4x\sqrt{-g}\,\left\{
(\na \ph)^2 + m_s^2\ph^2 + \xi R\ph^2 \right\}\,,
\label{scalar}
\eeq
where $(\na \ph)^2=g^{\mu\nu}\pa_\mu\ph\pa_\nu\ph$;
\beq
S_{1/2} \,=\,i \int d^4 x\sqrt{-g}\,\left\{\, 
{\bar \psi}\,[\ga^{\mu}\na_\mu + im_f]\,\psi\,\right\}\,;
\label{spinor}
\eeq
and
\beq
S_{1}\,=\,\int d^4 x\sqrt{-g}\,
\left\{\, - \frac14\,F_{\mu\nu}F^{\mu\nu}\,\right\}\,.
\label{vector}
\eeq

The massless matter fields possess local conformal invariance 
in the case where scalars couple to gravity in a nonminimal 
conformal way, with $\xi=1/6$. In what follows, we consider 
only this value, because it provides the traceless
stress tensor for the massless theory at the classical level. 
It is well known that at the one-loop level 
the condition $\xi=1/6$ holds under renormalization 
(see e.g. \cite{book}), but at higher loop this is not the 
case \cite{hath}. 
However, even at higher loops the deviation from $1/6$ can 
be very small \cite{brv}, such that $\xi=1/6$ can serve as
a good approximation.
Furthermore, massless fermions are conformal invariant, 
also the gauge fields can be treated as massless and 
conformal at very high energies.

The renormalization of the Einstein-Hilbert, cosmological
and $\sqrt{-g}R^2$ terms in (\ref{vacuum}) is absent 
in the case of conformal massless fields, and from this 
point of view their presence is not necessary. Indeed, we 
have to introduce the Einstein-Hilbert term because it 
is proved to fit with all experimental and observational 
data at the cosmic scale. For the same reason we introduce
the cosmological term which was detected in the 
recent observations \cite{sn}. Regarding the 
$\int \sqrt{-g} R^2$ term, we have to accept the presence 
of this term since it is important for the renormalizability
at higher loops. But its coefficient $a_4$ may be 
safely taken very small \cite{brv} such that it is negligible 
compared to the quantum corrections of the same form 
(see Eq. (\ref{quantum})). For this reason, in what follows 
we shall 
simply take $a_4=0$. Under this choice the action $S_{HD}$ 
satisfies the Noether identity
\beq
- \, \frac{2}{\sqrt{-g}}\,g_{\mu\nu}
\frac{\de S_{HD}}{\de g_{\mu\nu}}\, = \, 0\,,
\label{identity1}
\eeq
which is usually interpreted as zero trace for the 
stress tensor of vacuum $T^\mu_\mu=0$. At quantum level,
this identity is violated by the anomaly such that 
\cite{duff}
\beq
T\,\,=\,\,<T_\mu^\mu>\,\,=\,\,- \frac{2}{\sqrt{-g}}\,g_{\mu\nu}
\frac{\de {\bar \Ga}^{(1)}}{\de g_{\mu\nu}}\,\,=\,\,
- \,(wC^2 + bE + c{\nabla^2} R)\,.
\label{main equation}
\eeq
Here ${\bar \Ga}^{(1)}$ is the quantum correction to the 
classical vacuum action (\ref{vacuum}), the coefficients 
$\,w$, $\,b\,$ and $\,c\,$ are the $\beta$-functions for 
the parameters $\,a_1,a_2,a_3\,$ correspondingly
\beq
w \,=\, \frac{1}{(4\pi)^2}\,\Big(
\frac{N_0}{120} + \frac{N_{1/2}}{20} + \frac{N_1}{10} \Big)\,,
\label{w}
\eeq
\beq
b \,=\, -\,\frac{1}{(4\pi)^2}\,\Big( \frac{N_0}{360} 
+ \frac{11\,N_{1/2}}{360} + \frac{31\,N_1}{180}\Big)\,,
\label{b}
\eeq
\beq
c \,=\, \frac{1}{(4\pi)^2}\,\Big( \frac{N_0}{180} + \frac{N_{1/2}}{30}
- \frac{N_1}{10}\Big) \,.
\label{c}
\eeq
Some observation is in order. The direct application of the 
dimensional 
regularization \cite{duff} leads to the ambiguous result
for the coefficient of the $\nabla^2 R$-term in the anomaly. 
We accept the expression for $\,c\,$ which emerges from other 
regularizations such as point-splitting \cite{chr}
and $\ze$-regularization \cite{dow}. 

The general solution for the anomaly-induced effective action is 
\cite{rei,frts}
$$
\Ga_{ind} \,\,=\,\, S_c[{\bar g}_{\mu\nu}] \,+\,\,
\int d^4 x\sqrt{-{\bar g}}\,\{
\,w\si {\bar C}^2 + b\si ({\bar E}-\frac23 {\bar {\na}}^2
{\bar R}) + 2b\si{\bar \De}_4\si \,-
$$
\beq
-\, \frac{1}{12}\,(c+\frac23 b)[{\bar R}
- 6({\bar \na}\si)^2 - 6({\bar {\na}}^2 \si)]^2)\}\,,
\label{quantum1}
\eeq
where 
$$
g_{\mu\nu}={\bar g}_{\mu\nu}\cdot e^{2\si}
\,,\,\,\,\,\,\,\,\,\,\,\,\,\,\,
\sqrt{-g}R^2=\sqrt{-{\bar g}}[{\bar R}
- 6({\bar \na}\si)^2 - 6({\bar {\na}}^2 \si)]^2
$$ 
and
$$
\De_4 = \nabla^4 + 2\,R^{\mu\nu}\na_\mu\na_\nu - \frac23\,R{\nabla^2}
+ \frac13\,(\na^\mu R)\na_\mu
$$
is the fourth derivative, conformal operator acting on scalars.
$\,S_c[{\bar g}_{\mu\nu}]=S_c[g_{\mu\nu}]\,$ 
is the ``integration constant'' for the 
equation (\ref{main equation}). In general, there is no regular
method for deriving $S_c[{\bar g}_{\mu\nu}]$. Fortunately,
for some cosmological problems the form of this functional 
has no importance. Let us consider an isotropic 
and homogeneous metric 
\beq
g_{\mu\nu}={\bar g}_{\mu\nu}\cdot a^2(\eta)\,,
\label{flat metric}
\eeq 
where $\eta$ is conformal time and 
\beq
d{\bar s}^2={\bar g}_{\mu\nu}dx^\mu dx^\nu 
= d\eta^2 - \frac{dr^2}{1-kr^2}-r^2 d\Omega\,.
\label{flat1}
\eeq 
Then, the conformal functional $S_c[{\bar g}_{\mu\nu}]$ is
constant, for it does not depend on $a(\eta)$ and 
does not contribute to the equations of motion. In this 
particular case, the solution (\ref{quantum1}) is an exact
one-loop quantum correction. 
However, if one takes a more general background, for example 
\beq
g_{\mu\nu}={\bar g}_{\mu\nu}\cdot a^2(\eta) 
+ h_{\mu\nu}(\eta,{\vec r})\,,
\label{flat2}
\eeq 
where $h_{\mu\nu}$ is a metric perturbation. Then the 
functional $S_c[{\bar g}_{\mu\nu}]$ becomes relevant,
because
$$
S_c[g_{\mu\nu}]=S_c[{\bar g}_{\mu\nu}+a^{-2}(\eta) 
\cdot h_{\mu\nu}(\eta,{\vec r})]
$$
and 
the solution (\ref{quantum1}) is only an approximation to the
unknown general expression. In this and the next two sections 
we shall consider the metrics of the  form (\ref{flat1}) and
in the section 5 the metric (\ref{flat2}). 

The total action with the quantum corrections 
\beq
S_t=S_{vacuum}+\Ga_{ind} \,,
\label{massless}
\eeq
leads to the following equation for the metric 
(\ref{flat metric}):
\beq  
\frac{{\stackrel{....}{a}}}{a}
+\frac{{3\stackrel{.}{a}} {\stackrel{...}{a}}}{a^2}
+\frac{{\stackrel{..}{a}}^{2}}{a^{2}}
-\left( 5+\frac{4b}{c}\right) 
\frac{{\stackrel{..}{a}} {\stackrel{.}{a}}^{2}}{a^3}
-2k\left( 1+\frac{2b}{c}\right)
 \frac{{\stackrel{..}{a}}}{a^{3}}
-\frac{M_{P}^{2}}{8\pi c}
\left( \frac{{\stackrel{..}{a}}}{a}+
\frac{{\stackrel{.}{a}}^{2}}{a^{2}}
+\frac{k}{a^{2}}-\frac{2\Lambda }{3}\right)
\,=\,0\,,
\label{foe}
\eeq
where we used the standard notation for the Planck mass
$\,M_P^2=1/G\,$. The point over the variable designates the 
derivative with respect to the physical time $t$, the last 
is related to the conformal time $\eta$ by the relation
$dt=a(\eta)d\eta$. 

The equation (\ref{foe}) has remarkable particular 
solutions for all three cases $k=0,\pm 1$: 
\beq
k=0 \,\,\,\,\,\,\,\,\,\,\,\,\,\,\,\,\,\,\,\,\,\,\,\,\,\,\,
a(t) \,=\, a_0 \cdot \exp(Ht)
\label{flat solution}
\eeq
\beq
k=-1 \,\,\,\,\,\,\,\,\,\,\,\,\,\,\,\,\,\,\,\,\,\,\,\,\,\,\,
a(t) \,=\, \frac{1}{H}\,{\rm sinh}(Ht)\,,
\label{negative solution}
\eeq
\beq
k=+1 \,\,\,\,\,\,\,\,\,\,\,\,\,\,\,\,\,\,\,\,\,\,\,\,\,\,\,
a(t) \,=\, \frac{1}{H}\,{\rm cosh}(Ht)\,,
\label{positive solution}
\eeq
where
the Hubble constant $H$ has the form 
(we do not consider solutions with negative $H$)
\beq
H\,=\, \frac{M_P}{\sqrt{-32\pi b}} \cdot \left(\,1\pm \,
\sqrt{1+\frac{64\pi b}{3}\,\frac{\Lambda }{M_P^2} 
\,}\right)^{1/2}\,.
\label{H}
\eeq
For a negative $\La$, there is only a particular solution 
with positive sign in (\ref{H}). 

The solutions for the case of $\La=0$ have
been found by Starobinsky \cite{star} (see also \cite{mamo}).
According to (\ref{H}) and (\ref{b}), the two 
solutions with $\pm$ exist only for $\La \geq 0$. 
Let us remark that in the high energy region the 
sign of  $\La$ is positive. Let us advocate this point. 
In the nowadays universe $\La$ is also positive but very small,
due to the cancelation of the induced and vacuum energy 
densities (see, e.g. \cite{weinberg,nova} and also 
\cite{SantFeliu} for the pedagogical introduction to 
the formal aspects of the cosmological constant problem).

The induced cosmological constant appears due to the 
spontaneous symmetry breaking in the Standard Model, as a 
vacuum energy corresponding to the minima of the Higgs 
potential 
\beq
V_{cl}=-\frac{1}{2}m^{2}\phi^{2}+\frac{f}{8}\phi^{4}\,,  
\label{2a}
\eeq
and has the classical value
\beq
\Lambda _{ind}=<V_{cl}>=-\frac{m^{4}}{2f} < 0
\label{nnn6}
\eeq
with the magnitude about 55 orders greater than the observable 
cosmological constant. Obviously, the vacuum counterpart
must be positive in order the provide a cancelation 
at low energies. But, at high energies (temperatures)
above $300\,GeV$ the symmetry in the potential (\ref{2a}) gets 
restored such that the induced part (\ref{nnn6}) disappears. 
After that the value of the cosmological constant will be 
given by the vacuum term with the corresponding quantum 
corrections \cite{nova}. Similar increase of $\La$ may be
expected in any other phase transition which could occur
at high energies. In particular, the contribution to the 
cosmological constant may be expected from the supersymmetry
breaking in case this breaking is spontaneous. The last 
observation is that after inflation, when the matter 
content of the Universe  gets closer to the equilibrium state,
the high-temperature radiation dominates over the cosmological 
term \cite{bludman}. 

In what follows, except if this is indicated explicitly, we 
assume that the cosmological constant  $\La$ in the high 
energy inflationary epoch  satisfies the relation 
$0 < \La \ll M_P^2$. Then the expansion of 
(\ref{H}) in the small parameter $\La /M_P^2$ gives the 
two values 
\beq
H\,=\,\sqrt{\frac{\Lambda }{3}}\,\,\,\,\,\,\,\,\,\,\,\,
{\rm and}\,\,\,\,\,\,\,\,\,\,\,\,
H\,=\, \sqrt{- \frac{M^2_P}{16\pi b} - \frac{\Lambda }{3} }
\,,
\label{HH}
\eeq
where the first solution is exactly that we meet without
the quantum corrections. The error in both cases 
has the order of magnitude  $\,\,\,\La/M_P << H.\,\,$
In the present-day universe this means that what we 
observe is really $\La$, without significant quantum
corrections. In the inflationary epoch, if we compare the 
second solution (\ref{HH}) and the 
ones in the $\La=0$ case, the spectrum of $H$ is modified 
by the relatively small $\,\La$-term such that the two values 
of $\,H\,$ become closer to each other \cite{nova}. 

As we shall see in the next section, the inflationary 
solutions (\ref{flat solution}), (\ref{negative solution}) 
and (\ref{positive solution}) are stable for $c>0$ 
and unstable for $c<0$ \cite{star} (see also \cite{anju}).
According to (\ref{c}),  $c<0$ is equivalent to the
condition for a particle content of the theory 
\beq
N_1 \,>\,\frac13\,N_{1/2}\,+\,\frac{1}{18}\,N_{0}\,.
\label{condition}
\eeq

The unique way to provide the non-stability of 
inflation in the case of $c>0$ is to introduce 
some additional term or terms into the classical action
of vacuum. For instance, the $\int \sqrt{-g}R^2$-term 
with the sufficiently large positive coefficient does 
this job. But, for the first sight, this term is not 
necessary. 
Let us notice, that the quantum correction (\ref{quantum1})
is valid not only in the Early Universe but also at the 
later stages of the evolution. However, the content 
$N_0,N_{1/2},N_1$ may be different, because for the small
energies the loops of the fields with large masses decouple. 
Let us consider the present-day universe as an example. 
The value of the Hubble constant today is extremely small, 
about $\,H\sim 10^{-42}\,GeV$, that is about 30 orders 
smaller than the mass of the lightest neutrino. Therefore, 
the unique particle which makes a nontrivial contribution 
to the anomaly is the photon 
\footnote{If the neutrino would be massless, there 
could be an equality in (\ref{condition}).}.
Then, the relation (\ref{condition}) testifies a 
non-stability and the present-day universe is safe without 
the special $\,\int \sqrt{-g}R^2$-term. We shall continue
the discussion of this example in the section 3.

The condition (\ref{condition}) enables one to construct 
a very attractive inflationary scenario \cite{insusy}.
The universe could start in the stable phase, such that 
inflation starts independent on the initial data. The 
simplest way to provide stability in (\ref{condition})
is to assume supersymmetry in the high energy region and 
also suppose that the supersymmetry is absent at a
lower energy because the sparticles are heavy \cite{insusy}. 
If, in the course of inflation, 
the magnitude of the Hubble constant decreases, then at 
some point the loops of the sparticles decouple and the
matter content $N_{0,1/2,1}$ gets modified. As a result, 
the sign of the inequality (\ref{condition}) changes to 
the opposite and the universe falls into the non-stable 
inflation with the eventual transition to the FRW evolution. 
The important problem is that why the Hubble parameter 
should decrease during the inflation? Indeed, the 
Starobinsky solutions (\ref{flat solution}), 
(\ref{negative solution}) and (\ref{positive solution})
are characterized by the constant $H$. The 
answer to this question is that the value of $H$ really
decreases if we go beyond the massless approximation
\cite{shocom}. We shall consider the effect of massive 
fields below.

Another interesting aspect is the possible role of matter. 
Here we have two different situations. The first is the 
possibility to have a set of heavy massive fields at the 
beginning of inflation. These fields can be remnants of 
the string phase transition. For our model it is important 
that this massive content does not destroy the stability 
which holds at the beginning of the stable inflation. Let 
us suppose that these fields create zero pressure. Then 
the equation (\ref{foe}) is modified by adding the term 
$\,\rho_m a^{-3}\,$ to the {\it r.h.s.}. The investigation 
of the equation with this term and with $\La=k=0$ 
\beq  
\frac{{\stackrel{....}{a}}}{a}
+\frac{{3\stackrel{.}{a}} {\stackrel{...}{a}}}{a^2}
+\frac{{\stackrel{..}{a}}^{2}}{a^{2}}
-\left( 5+\frac{4b}{c}\right) 
\frac{{\stackrel{..}{a}} {\stackrel{.}{a}}^{2}}{a^3}
-2k\left( 1+\frac{2b}{c}\right)
 \frac{{\stackrel{..}{a}}}{a^{3}}
-\frac{M_{P}^{2}}{8\pi c}
\left( \frac{{\stackrel{..}{a}}}{a}+
\frac{{\stackrel{.}{a}}^{2}}{a^{2}}\right)
\,=\,-\,\frac{\rho_m}{c\,a^3}\,,
\label{matter}
\eeq
has been performed numerically. The result is such that even 
for $\rho_m = M_P^4$ the system does not loose stability and
the stabilization performs in a few Planck times. The plots 
are almost identical to the matterless case (see Fig. 2). The 
reason of this output is that the new term decreases very fast 
during the inflation and becomes irrelevant.

Second aspect is the transition to the FRW evolution after 
some of the massive fields decouple and the inflation 
becomes unstable. In order to understand the situation 
better, let us just substitute the corresponding FRW
solution $a(t)\sim t^{2/3}$ into the equation (\ref{matter}).
It is easy to see that the ``classical'' terms - 
both Einstein and matter, both behave like $1/t^2$, while
the quantum correction - with higher derivatives, behave
like $1/t^4$. Indeed, $H(t)\to 0$, as usual. Therefore, 
with the properly taken initial 
data for the unstable inflation, the FRW solution is 
a perfect approximation for the later epochs. 
In \cite{shocom} we have checked that the ``proper''
initial data emerge naturally in the framework of the 
sharp cut-off approach to the decoupling. 


\subsection{Massive case}

$\,\,\,\,\,\,\,$
Now, let us assume that some of the scalar and 
fermion fields are massive. At high energies, the leading 
vacuum quantum 
effect of the masses is the renormalization of the 
Newton and cosmological terms. In \cite{shocom} the 
renormalization of the Newton term has been taken into
account using some special approach which gives direct 
link to the cosmological applications. Here we shall follow 
the same method, but perform the calculations in a more 
detailed and general form. In particular, we shall take 
into account the renormalization of the cosmological term.

The conformal invariance 
of the actions (\ref{scalar}) and (\ref{spinor}) is
violated by the masses and we can not use the conformal 
anomaly to derive quantum corrections. But, this 
can be changed if we apply the conformal description 
of the massive theory in the framework of the cosmon model 
\cite{cosmon} (see also \cite{cosmon2} for other 
applications of the cosmon model). Let us replace the 
massive parameters by the powers of a new auxiliary scalar 
field $\,\chi$ according to  
$$
m_s^2 \to \frac{m_s^2}{M^2}\,\chi^2 
\,,\,\,\,\,\,\,\,\,\,\,\,\,\,\,\,\,\,\,\,\,\,\,\,\,\,\, 
m_f \to \frac{m_f}{M}\,\chi 
\,,
$$
\beq
\frac{1}{16\pi G}\,R \,\to\, \frac{M_P^2}{16\pi M^2}\,
\left[\,R\chi^2 + 6\,(\pa \chi)^2\,\right]
\,,\,\,\,\,\,\,\,\,\,\,\,\,\,\,\,\,\,\,\,\,\,\,\,\,\,\, 
\La \to \frac{\La}{M^2}\,\chi^2 \,,
\label{replace}
\eeq
where $\,M\,$ is some dimensional parameter. 
In the original version of the cosmon model \cite{cosmon},
the introduction of the new scalar fields $\chi$ is called 
to provide the dilatation invariance with the consequent 
breaking of the dilatation symmetry. In our case we need a 
more general, local conformal invariance. Therefore, we 
postulate that the auxiliary field $\chi$ transforms as
\begin{eqnarray}
\chi \to \chi\,e^{-\sigma}\,,\,\,\,\,\,\,\,\,\,\,\,\,\,\,\, 
\si=\si(x)\,,
\label{auxiliar}
\end{eqnarray}
independent of the space-time dimension 
\footnote{Other fields transform 
in a usual way: $\,g_{\mu\nu}\to g_{\mu\nu}\,\exp(2\si) 
\,,\,\, \varphi \to \varphi\,\exp(-\frac{n-2}{2}\si)
\,,\,\,\, \psi \to \psi\,\exp(-\frac{n-1}{2}\si).\,$
Here we use the dimensional regularization, because it 
helps to simplify the calculations.}.
After we apply the procedure (\ref{replace}), the massive 
terms in (\ref{scalar}) and (\ref{spinor}) 
are replaced by Yukawa and quartic scalar 
interactions between physical fermion and scalar fields
and the new auxiliary scalar $\,\chi$. 
The advantage of this procedure is that the new actions 
in both matter and gravitational sectors become conformal 
invariant. This invariance is violated by a conformal 
anomaly and one can derive the effective action of the 
background fields $g_{\mu\nu}, \chi$ using the
anomaly-induced effective action scheme.

The divergences of the theory in the conformal 
representation have the form 
\beq
\Ga^{(1)}_{div} = - \frac{\mu^{n-4}}{(n-4)}\,
\int d^nx\sqrt{-g}\,\Big\{ wC^2 + bE + c{\nabla^2}R 
+\,\frac{f}{M^2}\,[R\chi^2 + 6(\pa \chi)^2]
\,\,+\,\,\frac{g}{M^4}\,\chi^4\Big\}\,,
\label{div}
\eeq
where the coefficients $w,b$ and $c$ are given by Eqs. 
(\ref{w}), (\ref{b}) and (\ref{c}) correspondingly, and
we introduced useful notations for the dimensional 
quantities
\beq
f\,=\,\frac{1}{3(4\pi)^2}\,\sum_{f}\,{N_f\,m_f^2}\,,
\label{f}
\eeq
and
\beq
g\,=\,\frac{1}{2(4\pi)^2}\,\sum_{s} \,{N_s\,m_s^4}
-\frac{2}{(4\pi)^2}\sum_{f}\,{N_f\,m_f^4}\,.
\label{g}
\eeq
Here the sums are taken 
over all massive fermion $f$ and scalar $s$ fields with 
the masses $m_f$ and $m_s$ correspondingly. 
$N_f$ and $N_s$ are multiplicities of fermions and scalars,
for example $N_{lepton}=1$ and $N_{quark}=3$. In order 
to derive the conformal anomaly, one needs to consider
the classical Noether identity for the vacuum part of 
the effective action
\beq 
{\cal T}\, = \,-\,\frac{2}{\sqrt{-g}} g_{\mu\nu}
\,\frac{\de S_{vac}}{\de g_{\mu\nu}}
+  \frac{1}{\sqrt{-g}}\,\chi\,
\frac{\de S_{vac}}{\de \chi}\,= \,0\,.
\label{vacu}
\eeq
It is easy to see that, in the theory under discussion, the
conformal invariance does not mean that the stress tensor is 
traceless, 
but instead that the quantity $\,{\cal T}\,$ is zero. 
Correspondingly, the conformal anomaly means 
$\,<{\cal T}>\,\neq\,0$ instead of usual 
$\,<{T}^\mu_\mu>\,\neq\,0$ \cite{anhesh}. However, since both 
$\chi$-dependent terms in (\ref{div}) have the same conformal 
properties as the square of the Weyl tensor, they 
can be absorbed into the Weyl term such that the problem 
of deriving anomaly reduces to the standard one. Finally, we 
arrive at the expression 
\beq
<{\cal T}> \,=\, -\, \Big\{\,\,
wC^2\,+\,bE\,+\,c{\nabla^2}R \,+\,\frac{f}{M^2}
\,[R\chi^2+6(\pa \chi)^2]
\,+\,\frac{g}{M^4}\,\chi^4\,\,\Big\}\,.
\label{trace anomaly}
\eeq
In the standard way, using the parametrization 
$g_{\mu\nu}=e^{2\si}{\bar g}_{\mu\nu}$ and 
$\chi=e^{-\si}{\bar \chi}$, one can derive the 
anomaly-induced effective action of the background fields
$g_{\mu\nu}$ and $\chi$ 
$$
\Ga_{ind} \,=\, S_c[g_{\mu\nu}, \chi]
\,+\,\int d^4 x\sqrt{-{\bar g}} \,\{w{\bar C}^2\sigma
+ b({\bar E} -\frac23 {\bar \nabla}^2 {\bar R})\sigma
+ 2 b\,\sigma{\bar \Delta}\sigma + 
$$
\beq
+ \frac{f}{M^2}\,[\,{\bar R}{\bar \chi}^2 + 6\,(\partial {\bar
\chi})^2]\sigma\, + \, \frac{g}{M^4}\,{\bar \chi}^4\sigma\,\} 
\,\,-\,\, \frac{3c+2b}{36}\,\int d^4x\sqrt{-g}\,R^2\,.
\label{quantum}
\eeq

The last step is to fix the conformal unitary gauge
$\,\chi = {\bar \chi}\,e^{-\si}=M$, such that the 
classical Einstein-Hilbert and cosmological terms acquire 
their standard form. The one-loop effective action
becomes
$$
\Gamma \,\,=\,\,S_{vacuum}\,+\,\Ga_{ind} [\chi \to M] 
\,\,=\,\,S_{HD}+S_c[g_{\mu\nu}, M] \,+\,
$$
\vskip 1mm
$$
\,+\, 
\int d^4 x\sqrt{-{\bar g}} \,\{w{\bar C}^2\sigma
\,+\, b\,({\bar E} -\frac23 {\bar \nabla}^2 {\bar R})\,\sigma
\,+\, 2 b\,\sigma{\bar \Delta}\sigma \}
\,-\, \frac{3c+2b}{36}\,\int d^4x\sqrt{-g}\,R^2\,-
$$
\beq
-\,\int d^4 x\sqrt{-{\bar g}} 
\,e^{2\si}\,[{\bar R}+6({\bar \na}\si)^2]
\,\cdot\,\Big[\,
\frac{1}{16\pi G} - f\cdot\si\,\Big]
- \int d^4 x\sqrt{-{\bar g}}\,e^{4\si}\,\cdot\,
\Big[\frac{\La}{8\pi G}\,-\,g\cdot\si\,\Big] \,.
\label{quantum for massive}
\eeq
It is worth noticing that the expression 
(\ref{quantum for massive}) does not depend on the magnitude 
of $M$, but only on the fields and parameters of the theory. 
This effective action differs from the 
massless counterpart (\ref{massless}) in two respects. First, 
there are new terms which emerge due to the renormalization of the
Einstein-Hilbert and cosmological term. Second, the integration
constant $S_c[g_{\mu\nu}, M]$ is not irrelevant anymore, since 
we imposed the conformal unitary gauge. Therefore, 
the expression (\ref{quantum for massive}) can be regarded only 
as an approximation to the effective action of vacuum for the 
massive fields and we have to learn the limits of validity 
for this approximation.  

The higher-derivative part of the Eq. (\ref{quantum for massive}) 
is identical to that for the massless fields, as it has to be 
(see, e.g. \cite{birdav,mamo}). Furthermore, there is 
a strong link between the anomaly-induced effective
action Eq.(\ref{quantum for massive}) and the quantum 
corrections coming from the renormalization group. 
The expansion of the homogeneous, isotropic universe means 
a conformal transformation of the metric 
$\,g_{\mu\nu}(t) = {\bar g}_{\mu\nu} \exp\,[\si(\eta)]$.
On the other hand, the renormalization group in curved
space-time corresponds to the scale transformation of the 
metric $\,g_{\mu\nu} \to g_{\mu\nu}\cdot e^{-2\tau}\,$ 
simultaneously with the inverse transformation of all 
dimensional quantities \cite{book}. For any $\mu$ we 
have $\mu \to \mu\cdot e^{\tau}$. One can compare the 
$\,\sigma\,$-dependence of the anomaly-induced effective 
action (\ref{quantum for massive}) and the $\,\tau$-dependence 
of the renormalization-group improved classical action
\beq
S_{vac.imp}\,=\,S_{vacuum}[P(\tau)]\,,
\label{RGEA}
\eeq
where $P=\{a_{1,2,3,4},G,\La\}$ denote 
vacuum parameters of the theory and $P(\tau)=P_0 + \beta_P\tau$.
The expression (\ref{RGEA}) is a leading-log
approximation for the solution of the renormalization 
group equation for the effective action \cite{book}
\begin{eqnarray}
\Gamma[e^{-2\tau}g_{\alpha\beta},{\Phi_i},P,\mu ] =
\Gamma[g_{\alpha\beta},{\Phi_i}(\tau),P(\tau),\mu ]\,,
\label{RG}
\end{eqnarray}
where ${\Phi_i}$ are matter fields. 
It is easy to see that (\ref{quantum for massive}) becomes
completely equivalent to (\ref{RGEA}) if we set $\sigma=const$.
The coefficient $f$ is a factor of the $\beta$-function
for the inverse Newton constant $\,1/16\pi G$ and the 
coefficient $g$ is a factor of the $\beta$-function for the 
cosmological term $\La/8\pi G$. Indeed, (\ref{RG}) can be 
considered as a generalization of the renormalization group 
improved classical action (\ref{RGEA}).

Therefore, we have strong reasons to regard Eq. 
(\ref{quantum for massive}) as a leading-log approximation for 
the effective action for the massive fields.  This approximation 
picks up the logarithmic quantum corrections and is reliable 
in the high energy region where masses of the fields are much 
smaller than the Hubble constant $H$. On the other hand, at low 
energies the massive fields decouple and their quantum 
effects become negligible. In this paper we shall interpolate 
between these two regimes using the MS scheme and the
``sharp cut off'' approximation at some specific scale $M_*$. 
This scale is defined such that at $H =M_*$ a sufficient 
number of the 
sparticles decouple and the inequality (\ref{condition}) 
changes its sign.  Within our approximation, the functional 
$S_c$ contains the sub-leading non-log corrections and can 
be disregarded. Let us remark that the $M_*$ scale
can be different from the supersymmetry breaking scale
$M_{SUSY}$.  In particular, we can suppose that for a GUT 
model $M_{SUSY}$ is about $10^{16}\,GeV$ while the upper 
bound for $M_*$ is $10^{14}\,GeV$. This bound appears 
because the amplitude of the generated gravitational waves 
has an admissible range only for $H/M_P\leq 10^{-5}$ in the 
last stage ($\approx 65\,\,e$-folds) of inflation.  

The equation of motion for $\si(t)$ has the form 
$$
{\stackrel{....}{\sigma }}
+7{\stackrel{...}{\sigma }}{\stackrel{.}{\sigma }}
+4\,{{\stackrel{..}{\sigma }}}^{2}
+4\,\Big( 3-\,\frac{b}{c}\Big) 
\,{\stackrel{..}{\sigma }}{{\stackrel{.}{\sigma }}}^{2}
-4\,\frac{b}{c}\,{{\stackrel{.}{\sigma }}}^{4}
-2k\left( 1+\frac{2b}{c}\right) 
( {{\stackrel{.}{\sigma }}}^{2}
+{\stackrel{..}{\sigma }}) e^{-2\sigma }\,-
$$
\beq
- \,\frac{M_P^2}{8\pi c}\,\Big[
\,( {\stackrel{..}{\sigma}}\,+2{{\stackrel{.}{\sigma }}}^{2}
+ke^{-2\sigma }
)\cdot (1-\tilde{f}\si)-\frac12\,\tilde{f}
\dot{\sigma}^2\,\Big]\,\, + \,\, \frac{M_P^2\La}{12\pi c}
\,(1 - \tilde{g}\si-\tilde{g}/4)\,=\,0\,,
\label{central sigma}
\eeq
where we introduced useful notations
\beq
\tilde{f} = \frac{16\pi f}{M_P^2} = 
\frac{1}{3\pi}\,\sum_{f}\,\frac{N_f\,m_f^2}{M_P^2} 
\,;\,\,\,\,\,\,\,\,\,\,\,\,\,\,\, 
\tilde{g} = \frac{8\pi g}{M_P^2\La}
\,=\,\frac{1}{4\pi}\,\sum_{s}\,\frac{N_s\,m_s^4}{M_P^2\La}
\,-\,\frac{1}{\pi}\,\sum_{f}\,\frac{N_f\,m_f^4}{M_P^2\La}\,.
\label{replace11}
\eeq

It proves useful to rewrite this equation, also, in terms
of $a(t)\,\,$\footnote{We correct a small mistake in the 
similar expression (without the cosmological constant 
term) in \cite{shocom}. This correction does 
not modify the behavior of $a(t)$.}
$$  
\frac{{\stackrel{....}{a}}}{a}+3\,
\frac{\,{\stackrel{.}{a}}}{a}\,{\,}
\frac{{\stackrel{...}{a}}}{a}
+\frac{\,{\stackrel{..}{a}}^{2}}{a^{2}}
\,-\Big( 5+\frac{4b}{c}\Big)\, \frac{\,{\stackrel{..}{a}}}{a}
\,\frac{{\stackrel{.}{a}}^{2}}{a^{2}}
-2k\Big( 1+\frac{2b}{c}\Big)\frac{\,{\stackrel{..}{a}}}{a^{3}}\,-
$$
\beq
-\,\frac{M_P^2}{8\pi c}
\,\Big[\,\Big(\,\frac{{\stackrel{..}{a}}}{a}
+ \frac{\dot{a}^2}{a^2} + \frac{k}{a^{2}}
\Big)
\cdot (1-\tilde{f}\cdot{\rm ln} a)
\,-\,\frac{\tilde{f}}{2}\,\frac{\dot{a}^2}{a^2}\,\Big]
\,\, + \,\, \frac{M_P^2\La}{12\pi c}
\,(\,1 - \tilde{g}\cdot \ln a - \tilde{g}/4\,)\,=\,0\,.
\label{central a}
\eeq

The solution of the equation (\ref{central sigma}) can 
not be performed analytically in the general form, but 
it is obvious that the approximate solution for small 
$\,\,\ln a(t)$ can be obtained by the replacement
\beq
M_P^2 \to {\tilde M}_P^2 =
M_P^2\,\Big[1 - \tilde{f}\,\ln a(t)\Big]
\,;\,\,\,\,\,\,\,\,\,\,\,\,\,\,\,\,\,\,\,\,\, 
\La  \to {\tilde \La} = 
\La \Big\{1 - \tilde{g} \,\ln\,a(t)\Big\}
\label{replace1}
\eeq
in the expression for the Hubble parameter (\ref{H}) 
corresponding to the solutions (\ref{flat solution}),
(\ref{negative solution}) and (\ref{positive solution}).
Another possibility is the numerical integration of 
the equation (\ref{central sigma}). We present the 
plots produced by the numerical integration at 
Figure 1 for the case $\La=0$. These plots were obtained 
using Mathematica program \cite{m4} and also FORTRAN and 
standard algorithms of \cite{numerical}. 
It is easy to see that the initial parts of these numerical 
solutions are identical to the ones which were obtained 
in \cite{shocom}. Let us make one more step in 
the understanding of these plots and the quality 
of the approximation (\ref{replace1}). For the sake of 
simplicity, we shall consider only the value $\La=0$
\footnote{The numerical analysis shows that this is 
a good approximation for $\La \ll {\tilde f}\,M_P^2$.}. 
In this case one can use (\ref{replace1}) and 
easily integrate the equation 
\beq
{\dot \si}=H=H_0\sqrt{1 - \tilde{f}\si}
\,,\,\,\,\,\,\,\,\,\,
H_0=\frac{M_P}{\sqrt{-16b}}
\,,\,\,\,\,\,\,\,\,\,
\si(0)=0\,.
\label{simple}
\eeq
and find its solution in the form
\beq
\si(t)\,=\,H_0t\,-\,\frac{H^2_0}{4}\,\tilde{f}\,t^2\,.
\label{parabola}
\eeq
One might expect that this simple formula should serve as
a reasonable approximation only at the initial stage 
when $\si(t)=\ln a(t)$ is small. However, if we compare the 
plots (a) and (b) at the Figure 1 and the parabola Fig. 1c
(\ref{parabola}), the numerical difference is 
remarkably small - about $0.1$ at the maximal point 
with $\si_{max}=10^4$ (for the toy model with the MSSM 
field content and ${\tilde f}=10^{-5}$). The two solutions 
are almost identical for {\it all} $t$. 
Remarkably, this is not a specific feature 
of this particular value of $\tilde{f}$. If we substitute
(\ref{parabola}) into the equation (\ref{central sigma}),
the result is 
$$
-1.2 \cdot 10^{-8} 
+ (25.9 \,{\tilde f} - 2.59 \cdot 10^{-8}) \,t 
+ (3.14  \cdot 10^{-17} - 3.14  \cdot 10^{-8} f) \,t^2 +
$$$$
+ (-1.22  \cdot 10^{-26} + 1.22  \cdot 10^{-17} f) \,t^3 
+ (1.48  \cdot 10^{-36} - 1.48  \cdot 10^{-27} f) \,t^4
$$
that is indeed negligible for all situations of 
interest, when ${\tilde f}$ is very small.
This unexpected feature of the solution is 
extremely useful, for we can sometimes use simple
expression (\ref{parabola}) instead of the complicated
numerical solution of the equation (\ref{central sigma}).

In the last considerations we took into account
that ${\tilde f}$ and $\La$ are very small quantities.
This leads to the condition for the magnitude of the quantum 
corrections to the cosmological constant
\beq
\Big|
\frac12 \, \sum_{s}  m_s^4
-2\sum_{f} m_f^4\Big|\cdot \ln a(t)\ll 8\pi\La M_P^2\,.
\label{much less 2}
\eeq
Furthermore, one can 
request that the cosmological constant plays smaller 
role than the higher derivative induced contributions:
\beq
\Big|\frac12 \, \sum_{s}  m_s^4 -2\sum_{f} m_f^4\Big| 
\ll M_P^2\,\Big[1-\tilde{f}\ln a(t)\Big]^2
\,\,\,\,\,\,\,\,\,{\rm and}\,\,\,\,\,\,\,\,\,
\La \ll 8\pi M_P^2\,\Big[1-\tilde{f}\ln a(t)\Big]^2 \,.
\label{much less 3}
\eeq
Indeed, these conditions are satisfied at the beginning 
of inflation when $\ln a(t)$ is small and the higher 
derivative terms dominate over the cosmological constant
which must be, at most, of the order $M_{SUSY}^4$. It is 
convenient to assume the soft supersymmetry breaking 
scheme such that the magnitude of the cosmological constant
is much smaller than $M_*^4/M_P^2$.
When $\ln a(t)$ becomes large and 
the universe is close to the transition from stable to 
unstable inflation, we can not rely much on 
Eq.(\ref{replace1}) and have to use the numerical methods
to verify both stability of the tempered inflationary 
solution and the graceful exit to the FRW phase after
the supersymmetry breaks down.

\section{The conditions of stability in the massless case}

$\,\,\,\,\,\,\,$
In this section we shall consider the stability of the 
relatively simple massless case (\ref{foe}). Instead of the 
Eq. (\ref{foe}) one can take the third order $00$-component 
of the equation
\beq
R_{\mu\nu}-\frac12\,(R-2\La)\,g_{\mu\nu}\,=\,<T_{\mu\nu}>\,.
\label{correct}
\eeq 
Without the matter fields the $00$-equation is equivalent 
to the Eq. (\ref{foe}) which we used in the previous section
\cite{fhh,star,hhr}. The 
$00$-equation can be solved analytically \cite{star}
and the phase diagram shows the existence of the single 
attractor corresponding to the inflationary solution
in the stable case and various attractors, including 
$H=0$ in the non-stable case with $k=\La=0$. 
Unfortunately, our equations (\ref{central sigma}) or 
(\ref{central a}) are much more complicated than in the 
massless case and there it is not 
possible to solve them exactly. However, there is no
real need to do that. At the beginning of inflation 
the stabilization of
the exponential solution performs very fast \cite{anju}. 
Therefore, at the latest stages it is sufficient to check
the  asymptotic stability of the inflationary solution 
with respect to the small perturbations in the sense of 
Lyapunov (see, e.g. \cite{marot} regarding the stability 
issues in higher derivative models).

In this section we establish some important general 
properties of the stability under small perturbations
and will apply this knowledge for the approximate 
analytical and numerical analysis of the massive case
in the next section.

With respect to the criterion (\ref{condition}) of 
stability \cite{star,anju}, there are three relevant 
questions: i) which kind of perturbations one has to 
consider; ii) whether the choice of $k=0,1,-1$ may affect 
(\ref{condition}); iii) whether the presence of the 
cosmological constant may affect this criterion.

In order to address the issue i), let us  
make small perturbations in the equation (\ref{foe})
and set, for the sake of simplicity, $\La=0$ and $k=0$. 
First we try the perturbations of the form
$$
a(t)=a_0(t)+x(t)\,,\,\,\,\,\,\,\,\,\,\,\,\,\, 
{\rm where} \,\,\,\,\,\,\,\,\,\,\,\,\, 
a_0(t)=\exp(Ht)\,\,\,\,\,\,\,\,\,\,\,\,\, 
{\rm and} 
\,\,\,\,\,\,\,\,\,\,\,\,\, H=\frac{M_P}{\sqrt{-16\pi b}}\,.
$$
After simple calculations we arrive at the equation
for the perturbations $x(t)$
\beq
{\stackrel{....}{x}}+3H{\stackrel{...}{x}}
-\Big(3+\frac{2b}{c}\Big)\,H^2{\stackrel{..}{x}}
-\Big(7+\frac{4b}{c}\Big)\,H^3{\stackrel{.}{x}}
+6\Big(1+\frac{b}{c}\Big)\,H^4x=0\,.
\label{perturbations a}
\eeq
Indeed, the factors of $H$ can be easily removed by 
the simple rescaling of time, and then the standard
Routh-Hurwitz conditions (see the Appendix) show that 
the stability {\it can not be achieved} for any values 
of $b/c$. Of course, this contradicts our previous statement 
(\ref{condition}), and we can conclude that the 
perturbations of $a(t)$ do not serve for our purposes.

Fortunately, from the theory of ordinary differential 
equations (see, e.g. \cite{cezare})
we know that the stability may depend on the choice of
the variables. Therefore, let us consider the same problem 
in terms of $\si(t)$. The corresponding equation can be 
obtained from (\ref{central sigma}) by setting 
$\,\, \La=k={\tilde f}={\tilde g}=0 \,\,$ 
and making small perturbation of the form
$$
\si(t)=\si_0(t)+y(t)\,,\,\,\,\,\,\,\,\,\,\,\,\,\, 
{\rm where} \,\,\,\,\,\,\,\,\,\,\,\,\, 
\si_0(t)=Ht\,.
$$
The equation for $\,y(t)\,$ has the form
\beq
{\stackrel{....}{y}}+7H{\stackrel{...}{y}}
+ \left[4\Big(3-\frac{b}{c}\Big)\,H^2 
- \frac{M^2_P}{8\pi c}\right]{\stackrel{..}{y}}
- \left[\frac{16b}{c}\,H^3 
+ \frac{M^2_PH}{2\pi c}\right]
{\stackrel{.}{y}}\,=\,0\,.
\label{perturbations 1}
\eeq
or, taking the solution for $H$ into account, 
\beq
{\stackrel{....}{y}}+7H{\stackrel{...}{y}}
+2\Big(6-\frac{b}{c}\Big)\,H^2{\stackrel{..}{y}}
-\frac{8b}{c}H^3{\stackrel{.}{y}}\,=\,0\,.
\label{perturbations sigma}
\eeq
It is easy to see that in this case the stability is 
achieved when $c$ is positive that immediately gives 
(\ref{condition})
\footnote{Of course, the conditions of 
stability for $H(t)$ are exactly those for $\si(t)$.}.
The illusory contradiction between
the two results can be clarified if we identify 
$$
y(t)\,=\,\ln\left[\,1+\frac{x(t)}{a_0(t)}\,\right]\,.
$$
Since $a_0(t)$ grows up very fast, the growing 
modes in $x(t)$ may be negligible because they are 
suppressed by the factor of $1/a_0(t)$. Of course,
this is the case when $c>0$, while for $c<0$ there
is a mode in $x(t)$ which is growing up faster than
$a_0(t)$. Finally, we learned a very important lesson
for the consequent study of the complicated massive case: 
one has to perturb $\si(t)$, but not $a(t)$. The calculations 
for the $k\neq 0$ and $\La\neq 0$ cases are a bit more 
complicated, and we will not bother the reader with 
the details. The condition of stability is always the
same as in the simplest situation described above. But, 
it is important that this condition {\it must be} the 
same for $k=0,1,-1$ and {\it must be} independent on $\La$.
Let us formulate two simple theorems:
\vskip 3mm

{\it Theorem 1.} The value of $\La$ does
not influence the asymptotic stability. 
\vskip 1mm

{\it Proof}. The equation for the perturbation
$x(t)$ or $y(t)$ does not change if the original equation 
for $\,a(t)\,$ or $\,\si(t)\,$ is multiplied by the factor 
of $\,a^n\,$
or $\,\exp(n\si)$. Therefore, we can always isolate the 
cosmological term such that its contribution to the 
perturbation equation is zero. The dependence on $\La$ 
can be also seen in $\,H$, but the magnitude of $\,H\,$ 
is irrelevant for stability, because we can always 
renormalize time such that in the new equation $\,H\to 1$.
\vskip 1mm

{\it Observation}. This theorem is valid only for 
$\La \ll M_P^2$. 
\vskip 3mm

{\it Theorem 2.} The choice of $k=0,1,-1$ does
not influence the asymptotic stability and the
condition (\ref{condition}) is valid for all three 
cases. 
\vskip 1mm

{\it Proof}. It is sufficient to apply the corresponding 
theorem about the asymptotic stability in the limit 
$\,t\to\infty$ (see Appendix). 
In the equation for the perturbations one 
can always replace the coefficient functions by their 
asymptotic values. Since both $sinh(Ht)$ and $cosh(Ht)$ 
behave like an exponential at the late times, the conditions 
of stability for all three cases are the same. 
\vskip 1mm

{\it Illustration}. The equation for the perturbation
in the general massless case is 

\begin{eqnarray*}
&& \stackrel{....}{y}[t]+7\stackrel{.}{\si}_0
\stackrel{...}{y}[t]+\left[
\left( 1+\frac{2b}c\right) 
\left(\stackrel{.}{\si }_0^2+\stackrel{..}{\si}_0\right) 
+\frac{M_P^2}{16\pi c}\right] 4ke^{-2\si_0}h[t] \\
&&+\left[ 8\stackrel{..}{\si}_0
+4\left( 3-\frac bc \right) \stackrel{.}{\si}_0^2
-2k\left( 1+\frac{2b}c\right) e^{-2\si_0}
-\frac{M_P^2}{8\pi c} \right] \stackrel{..}{y}[t] \\
&&+\left[ 7\stackrel{...}{\si}_0
+8\left( 3-\frac bc \right) \stackrel{..}{\si}_0\stackrel{.}{\si}_0
-\frac{16b}c\stackrel{.}{\sigma}_0^3
-4k\left(1+\frac{2b}c\right)e^{-2\si_0}\stackrel{.}{\si}_0
- \frac{M_P^2}{2\pi c}\stackrel{.}{\si}_0\right] 
\stackrel{.}{y}[t]=0\,.
\end{eqnarray*}
If we neglect those terms in the coefficients of the 
perturbations which tend to zero at $t\to \infty$, the 
$k$ - dependence disappears. 

Thus, we have learned two lessons from the analysis 
of the massless case. It is sufficient to consider the case 
$\La=0$ and $k=0$, for this does not modify the stability of the 
solution. These statements can be used in the more realistic 
theory of inflation based on the effective action for the 
massive fields, but with a proper caution. In fact, the 
negligible role of $\La$ depends on that it is a constant. 
As we shall see in the next section, this is not true for
the effective action of massive fields.

Until now, we have considered only the stable inflationary 
solution with $c>0$. In other words, we were interested in
the high energy region where the supersymmetry is unbroken.
Obviously, another end of the energy scale also represents 
a great interest, for it enables one to perform a simple and
efficient test of the model. Let us consider, again, a nowaday
universe with $H=H_0 \sim 10^{-42}GeV$. According to the 
recent data \cite{sn}, this magnitude of the Hubble constant 
is mainly due to the contribution of the cosmological constant.
As we have already mentioned in the last section, the 
value of $c$ in the present-day universe is negative due 
to the contribution of the single massless particle - photon.
Therefore, we have to check whether the higher derivative terms
do not destroy the stability of the first solution in Eq. 
(\ref{HH}). Using (\ref{perturbations 1}), we arrive at the 
following characteristic equation for the perturbation of 
$\,H \to H + const \,\cdot\,\exp(\la t)$.
\beq
\la^3+7H\la^2 + \left[4\Big(3-\frac{b}{c}\Big)\,H^2 
- \frac{M^2_P}{8\pi c}\right]\la
- \left[\,\frac{16b}{c}\,H^3 
+ \frac{M^2_P}{2\pi c}\,H\,\right]\,=\,0\,.
\label{perturbations 2}
\eeq
Since the explicit solution of this equation is tedious, 
let us start from the $\La=0$ case. Then $H=0$ and the roots 
of the equation (\ref{perturbations 2}) are 
\beq
\la^{(0)}_1=0\,,\,\,\,\,\,\,\,\,\,\,\,\,\,\,\,\,\,
\la^{(0)}_{2/3}\,=\,\pm\,\frac{M_P}{\sqrt{8\pi |c|}}\,i\,.
\label{root 1}
\eeq
At this level, we do not have definite 
answer to our question about stability. Now, let us 
look for the solution of Eq. (\ref{perturbations 2})
making perturbations in (\ref{root 1}). Due to the 
huge difference between $H$ and $M_P$, this approximation
is perfect. The result is 
\beq
\la_1=-4H\,,\,\,\,\,\,\,\,\,\,\,\,\,\,\,\,\,\,
\la_{2/3}\,
=\,-\frac32\,H\,\pm\,\frac{M_P}{\sqrt{8\pi |c|}}\,i
\,,\,\,\,\,\,\,\,\,\,\,\,\,\,\,\,\,\,
{\rm where} \,\,\,\,\,\,\,\,\,\,\,\,\,\,\,\,\,
H=\sqrt{\frac{\La}{3}} > 0\,.
\label{root 2}
\eeq
It is very nice to see that the cosmological constant 
$\La>0$ really stabilizes the solution in the low-energy 
region, exactly as we should optimistically expect. 
The stability can be verified also for the earlier 
post-inflationary epoch, when the energy density of 
vacuum played smaller role than the density of 
radiation \cite{bludman,weinberg}. For the later epoch,
as we have already 
noticed, when we substitute the FRW solution 
into Eq. (\ref{foe}), the ``quantum'' terms decrease 
as $1/a^2$ compared to the Einstein-Hilbert and 
matter terms. Therefore, for the later epoch of the 
expanding post-inflationary universe the FRW is a very
good approximation. 


\section{The stability in the interpolation regime}

$\,\,\,\,\,\,\,$
At the very beginning of the inflationary period, the 
evolution goes exactly as in the massless case and the 
condition of stability must be the same (\ref{condition}). 
Later on, when $\,\si(t)\,$ becomes large enough, the 
deviation from the exponential inflation becomes greater.
Finally, the inflation slows down such that the Hubble 
parameter achieves the scale $M_*$, and the massive 
sparticles decouple. Then 
the condition (\ref{condition}) breaks down. Our purpose 
is to see whether the Eq. (\ref{condition}) remains the 
condition of stability for the modified equations 
(\ref{central sigma}) until the point $H \approx M_*$.

Before starting the numerical analysis, let us perform  
some analytic investigation of the stability in a useful 
approximation framework. It is sufficient to investigate  
the behavior of the perturbations in the physically 
important region corresponding to the last 65 $e$-folds
before $H$ achieves the value $M_*$. Looking at the plot 
at Fig. 1, we see that the dependence $\si(t)$ is 
rather smooth. Hence, we can divide this region into 
small intervals $(t_i,t_i+\De t_i)$ and consider 
$H(t)=H_i=const$ inside each of the intervals. Indeed,
$H$ will change from one interval to another, and this
may be the source of new perturbations. Of course, 
all this consideration is done for the stable 
case $c>0$, while for the non-stable case this approach
has no sense. For the purpose of analytic study, we 
accept the approximation (\ref{replace1}). 
Then, the direct calculations give the 
following equation for the perturbations 
$\si \to \si + y(t)$:
\beq
b_0 {\stackrel{....}{y}}+b_1 {\stackrel{...}{y}}
+ b_2 {\stackrel{..}{y}}
+ b_3 {\stackrel{.}{y}} + b_4 y \,=\, 0 \,,
\label{equa-pert}
\eeq
where
$$
b_0=1\,,\,\,\,\,\,\,\,\,\,\,\,\,\,
b_1=7{\tilde H}\,,\,\,\,\,\,\,\,\,\,\,\,\,\,
b_2=4\Big(3-\frac{b}{c}\Big){\tilde H}^2
- \frac{{\tilde M}_P^2}{8\pi c}\,,
$$
\beq
b_3=-\frac{16\,b}{c}\,{\tilde H}^3
- \frac{{\tilde M}_P^2 {\tilde H}}{2\pi c}
\,,\,\,\,\,\,\,\,\,\,\,\,\,\,
b_4=\frac{{\tilde M}_P^2}{4\pi c}\,\Big(
{\tilde H}^2{\tilde f}-\frac13\,{\tilde \La}{\tilde g}\Big)
\label{coefficients}
\eeq
and
\beq
{\tilde H}^2 = -\frac{{\tilde M}_P}{32\pi b}\,
\Big[\,1\,+\, \Big( 1+\frac{64\pi b}{3}\,
\frac{{\tilde \Lambda}}{{\tilde M}_P^2}\Big)^{1/2}\Big]\,.
\label{tilde}
\eeq
If we treat the slowly-varying $\si(t)$ as a constant, then
the stability depends on the signs of the Routh-Hurwitz 
determinants (see Appendix). Obviously, $D_1=b_1$ is 
positive. After some algebra we arrive at the expression
\beq
D_2=-\frac{21{\tilde M}_P^2 {\tilde H}}{8\pi b}\,\left[\,
1 + \Big(1- \frac{b}{7c}\Big) \cdot \Big( 1+\frac{64\pi b}{3}\,
\frac{{\tilde \Lambda}}{{\tilde M}_P^2}\Big)^{1/2}\right] > 0\,.
\label{D2}
\eeq
The general expression for $D_3$ is 
$$
D_3=-\frac{21{\tilde M}_P^4 {\tilde H}^2}{16\pi^2 bc}\,\left[\,
1 + \frac{64\pi b}{3}\,\frac{{\tilde \Lambda}}{{\tilde M}_P^2}
+ \Big(1- \frac{b}{7c}\Big) \cdot 
\Big( 1+\frac{64\pi b}{3}\,
\frac{{\tilde \Lambda}}{{\tilde M}_P^2}\Big)^{1/2}\right]-
$$
\beq
-\frac{49 {\tilde M}_P^2 {\tilde H}^2}{4\pi c}\,
\Big[ {\tilde H}^2 {\tilde f} 
- \frac13\,{\tilde \La} {\tilde g}\Big]\,.
\label{D3}
\eeq
Here only the last term may be negative. Let us notice that 
the values of ${\tilde f}$ and ${\tilde g}$ are small,
because they are proportional to the ratios of the particle
masses and the Planck mass (\ref{replace11}). Hence, we can 
safely conclude that $D_3$ is positive if the cosmological 
constant is not too large. Furthermore, $D_4=b_4\cdot D_3$ 
and the stability of the tempered inflation, within our 
framework, depends on the sign of the same expression which 
we met in (\ref{D3})
\beq
{\tilde H}^2{\tilde f}-\frac13\,{\tilde \La}{\tilde g}\,. 
\label{key}
\eeq
The positivity of the last expression is a criterion of 
stability in the last stage of the interpolation regime
- just before the decoupling of sparticles and consequent 
transition to the unstable regime. The sign of 
(\ref{key}) depends on the relation between the masses of 
the fermion and boson components and on the way of the 
supersymmetry breaking. The last defines the magnitude of 
the cosmological constant. If the supersymmetry breaking is 
spontaneous, the vacuum energy density 
$\,{\tilde \La}{\tilde M}_P^2$  may be as large as 
$M_{SUSY}^4$, that could be, in principle, much greater 
than $M_*^4$. Then, in the case of a positive value of 
${\tilde g}$ in (\ref{replace11}), the inflation would be 
destabilized by the cosmological constant earlier than at 
the $M_*$ scale. For the GUT-scale supersymmetry breaking 
this might be bad, because then the magnitude of the Hubble 
parameter in the last $65$ $e$-folds is too big
\footnote{For the lower scale supersymmetry breaking
this does not pose a problem.}. Thus, for the sake of 
simplicity we shall assume that ${\tilde g}$ is negative 
or that the supersymmetry breaking is not spontaneous such 
that the vacuum energy density does not exceed $M_*^4$. 
In each of 
these two cases the transition from the stable to unstable
inflation performs, qualitatively, in the same way as for
${\tilde \La}=0$ and the stability of inflation holds 
until the supersymmetry breaking at the $M_*$ scale.  

In the same framework we can investigate the 
oscillations of the Hubble parameter during inflation.
This is potentially important issue, since it is related to 
the possibility of a reheating. Let us look again at the
equation for the perturbations (\ref{equa-pert}). Typically, 
the characteristic equation (A4)   
has one or two couples of complex conjugate solutions
\beq
\la \, = \, - \, \al \, \pm \, i\be\,,
\label{complex}
\eeq
where $\al$ are positive in the stable case. Also, there 
may be real solutions. For example, in the case of the MSSM,
$\La=0$ and ${\tilde f}=10^{-5}$, we meet one 
couple of complex solutions
$\,\la_{1/2}\approx -1.46\pm  3.47i=\al \pm i\be\,$
and two real solutions $\la_3\approx -3.88$ and 
$\la_4 \approx -4.85\cdot 10^{-6}$. 
The solution for the perturbations is 
\beq
y(t) = C_0\,e^{-\al t}\,\cos(\be t + \ph)
+ C_1\,e^{-\la_3 t}\,+ C_2\,e^{-\la_4 t}\,.
\label{perturbated}
\eeq
We will be interested in the effect of $\be$. Of 
course, for the constant coefficient
$\,b_k\,$ in (\ref{coefficients}) the oscillations in 
the solution are invisible because they are suppressed by
the exponentials. However, let us remember that the 
coefficients $b_k$ are not constants, for they depend 
on ${\tilde H}={\tilde H}(t)$. We can consider ${\tilde H}$ 
as a constant only within a small interval  
$(t_i,t_i+\De t_i)$. When we pass from one such interval 
to another, the value of  ${\tilde H}$ changes. The
evolution of ${\tilde H}$ will result in the 
oscillations (\ref{perturbated}) and they may become
significant when ${\tilde H}$ changes faster. Let 
us evaluate this effect. If we define 
$\,{\bar b}_k = {\tilde H}^{-i} b_k\,$ and 
$\,{\bar \la} = - {\bar \al} + i{\bar \be}\,,\,$
$\, - {\bar \al}_{1/2}\,$
to be the solutions of the equation 
\beq
{\bar b}_0 {\bar \lambda}^4 
+ {\bar b}_1 {\bar \lambda}^{3} 
+ {\bar b}_2 {\bar \lambda}^{2} 
+ ... + {\bar b}_{3} {\bar \lambda} + {\bar b_4}\, = \,0\,.
\label{tilde1}
\eeq
Then ${\bar \al}_{1/2}$, ${\bar \al}$  and ${\bar \be}$ 
are constants and moreover 
\beq
\al_{1/2} = {\tilde H}{\bar \al}_{1/2} 
\cong H (1-{\tilde f}\si) {\bar \al}_{1/2}
\,,\,\,\,\,\,\,\,\,\,\,\,\,
\be = {\tilde H}{\bar \be} \cong H (1-{\tilde f}\si)
{\bar \be}
\,,\,\,\,\,\,\,\,\,\,\,\,\,
\al = {\tilde H}{\bar \be}
\cong H (1-{\tilde f}\si){\bar \al}\,.
\eeq
The last formula shows that the frequency of the
oscillations of $\si(t)$ and $H(t)$ is greatly 
decreasing during inflation. But, the same concerns 
also the damping parameter $\al$. 
Now, let us look at the plot  Fig. 1a. 
We can divide it into three regions 
with distinct properties: 1) The initial stage when
masses of the particles is irrelevant. Then, the 
$H(t)$ is almost constant and the perturbations 
are not visible. 2) The intermediate stage of the
"tempered" inflation with a significant effect of
the particle masses. In this region $H(t)$ changes
faster and one can expect oscillations with 
a relevant amplitude. It is easy to see that the 
amplitude of the oscillations of $H(t)$ has the
order of magnitude 
$\,\,\sqrt{\al^2+\be^2}\,C_0$.
3) The final region when $\si(t)$ is approaching 
a plateau and $H(t) \to 0$, where our approximation related 
to the ``sharp cut-off'' decoupling obviously fails.
In this region the sparticles start to decouple and the 
universe performs the transition to the non-stable inflation 
and eventually into the FRW-like evolution.


The numerical analysis of stability has been performed using the 
numerical simulations method, and we found that all curves rapidly
converge to the non-perturbed one in both the first and the last 
65 $e$-folds of inflation. We have chosen the initial data 
$\,{\stackrel{...}{\sigma }}(t_i)$,\, 
${\stackrel{..}{\sigma }}(t_i)$, \,
${\stackrel{.}{\sigma }}(t_i)$,\,
${\sigma}(t_i)\,$
for the equation (\ref{central sigma}) using the 
generator of random numbers independent for each 
derivative and then integrated 
this equation numerically. The choice of $t_i$
has been done in such a way that the corresponding 
$\si_i=\si(t_i)=\si_f-65$, where $\si_f$ corresponds
to $H=M_*$. In all cases we took the same initial 
value ${\sigma}(t_i)=1$. The measure of the deviation 
from the non-perturbed solution $H_{np}(t)$ can be  
taken, for example, in the form
$$
D=\left(\frac{{\stackrel{.}{H}} - {\stackrel{.}{H}}_{np}}
{H_{np}^2}\right)^2     
+\left(\frac{{\stackrel{..}{H}} - {\stackrel{..}{H}}_{np}}
{{H_{np}^3}}\right)^2
+\left(\frac{{\stackrel{...}{H}} - {\stackrel{...}{H}}_{np}}
{H_{np}^4}\right)^2\,.
$$
The result of the time-consuming
calculations is that, independent on the choice of 
the measure for the deviation, the solution $\si_0$
presented at Fig. 1 is stable. Some plots, illustrating the 
behavior of the perturbed solutions are presented 
at the Fig. 2. 
The plot of $H(t)$ for the last $65$ $e$-folds at Fig. 3
enables  one to see the oscillations \footnote{Similar 
plot with oscillations has been obtained 
using Mathematica program in the course of preparation
of the previous article \cite{shocom}. This plot has
not been included into \cite{shocom} due to the size
limits.} which are not visible 
in the scale of Fig. 1

Finally, the numerical solutions and approximate analytic 
investigation of the stability issues in this model completely 
agree between each other. We can observe, at Figure 2, an 
illustration of the stability of the inflationary solution, 
and at Figure 3 the oscillations of $H$. We can conclude, 
that our approximate analytic approach is sufficiently robust 
and is applicable to other problems, like the analysis
of the gravitational waves and (hopefully) of the density 
perturbations.

\section{On the amplitude of the gravitational waves}

$\,\,\,\,\,\,\,$
The most difficult test for a high derivative cosmological 
model is the behavior of the cosmological 
perturbations. In this section we shall present the 
preliminary results concerning metric perturbations 
(gravitational waves) during the last 65 $e$-folds of the
anomaly-induced inflation. Similar problem
was previously studied in \cite{star1,wave}.
In both papers, the analysis was restricted to the anomaly
produced by the massless quantum fields, and the small
difference in the equations for the perturbations is 
(in our terms) due to the different choice of the conformal 
invariant functional $S_c[g_{\mu\nu}]$ in Eq. (\ref{quantum1}). 
In the Ref. \cite{wave}, the derivation of equations for
the perturbations has been based on the anomaly-induced 
effective action and, in principle, this derivation can
be generalized for the effective actions Eq. (\ref{quantum})
and (\ref{quantum for massive}). However, this requires
a formulation
of the local covariant version of the effective actions
(\ref{quantum}) and (\ref{quantum for massive}), considerable
calculations of perturbations for a variable $H(t)$ and 
consequent numerical analysis. 
At the same time, there is a possibility to evaluate
the stability under the metric perturbations in a more 
economic way, and that is what we are going to discuss here. 
The detailed study of the gravitational waves will be 
performed separately.

Let us make some helpful observations about what we really 
need to know about the metric perturbations. It is well
known, that the only phenomenologically important region 
is the last 65 $e$-folds of inflation. All the gravitational 
waves which were generated before that are not visible 
in the present-day universe and hence they have no great 
importance. 
We shall follow the same procedure which proved reliable in 
the last  section when we investigated the perturbations of 
$\sigma(t)$. Thus, we divide the time interval corresponding 
to the last $65$ $e$-folds into the small pieces 
$(t_i,t_i+\De t_i)$. Within each interval $(t_i,t_i+\De t_i)$, 
the time derivatives of the Hubble parameter are negligible 
compared to its own magnitude, and we can safely take
constant $H$. In fact, this means that the last 
65 $e$-folds can be approximated by the massless model,
but with a greatly reduced magnitude of the Hubble 
constant and with much greater magnitude of $\si=\ln a\,\,$
\footnote{At this instant, we have to admit that the 
choice of the vacuum for the perturbations can be done
in the same way for the effective actions generated by
both massless and massive fields. It would be quite 
interesting to check its validity using some other approach 
to the derivation of the effective action for massive 
fields.}.
As we shall see below, the last point leads to the
dramatic difference with the original Starobinsky model
\cite{star1}.

For the sake of simplicity, let us present the calculations 
for the $\La=0$ case only. 
Using the results of \cite{wave}, we arrive at the following 
equations for the perturbations in the massive case
$$
  b_0\stackrel{....}{h}
+ b_1\stackrel{...}{h}
+ b_2\stackrel{..}{h}
+ b_3\stackrel{.}{h}
+ b_4h+
$$
\beq
+n_1e^{-2\si}\na^2\stackrel{.}{h}
+n_2e^{-2\si}\na^2\stackrel{..}{h}
+n_3e^{-4\si}\na^4h\,=\,0\,,
\label{wave}
\eeq
where $h \equiv h(t,{\vec x})$ and
$$
b_0 = a_1+w\cdot\si(t)\,,
\,\,\,\,\,\,\,\,\,\,\,\,\,\,\,\,\,\,\,\,\,\,
b_1 = 6H[a_1+w\cdot\si(t)]+2wH\,,
$$
$$
b_2 = 11H^2[a_1+w\cdot\si(t)]+H^2(c-b/2+7w)\,,
$$
\beq
b_3 = 6H^3[a_1+w\cdot\si(t)]+H^3(3c-3b/2+5w)\,,
\,\,\,\,\,\,\,\,\,\,\,\,\,\,\,\,\,\,\,\,\,\,
b_4 = -12H^4b\,,
\label{beta}
\eeq
\beq
n_1 = -2H[a_1+w+w\cdot\si(t)]\,,
\,\,\,\,\,\,\,\,\,\,\,\,\,\,\,\,\,\,\,\,\,\,
n_2 = -2[a_1w\cdot\si(t)]\,,
\,\,\,\,\,\,\,\,\,\,\,\,\,\,\,\,\,\,\,\,\,\,
n_3 = a_1+w\cdot\si(t)\,.
\label{n}
\eeq
We are mainly interested in the dynamics of the 
perturbations in the last 65 $e$-folds of inflation 
when the magnitude of $\si$ is at least of the order 
$10^{4}$. Then, the first observation regarding Eq. (\ref{wave}) 
is that the terms with the space derivatives are suppressed 
by the factor of $e^{2\si}$ and can be safely neglected
such that the equation does not depend on the magnitude
$\,n\,$ of the wavenumber. 
This does not guarantee that the spectrum is flat,
because the dependence on $n$ may appear from the 
initial data. For example, in the case of the 
perturbations of quantum origin the initial data 
are given by \cite{wave}
\begin{equation}
h_0 \propto \frac{1}{\sqrt{2n}} \quad , \quad \dot h_0 \propto
\sqrt{\frac{n}{2}} \quad,
\quad \ddot h_0 \propto \frac{n^{3/2}}{\sqrt{2}} \quad , \quad
{\stackrel{...} h}_0
\propto
\frac{n^{5/2}}{\sqrt{2}} \quad .
\label{initi}
\end{equation}

Even without the $\,n$-dependent terms, the equation 
(\ref{wave}) is still complicated, but its analysis can be 
essentially simplified if we remember that $\si \gg 1$. Taking 
into account that the $\be$-functions (\ref{w}), (\ref{b}),
(\ref{c}) all have the same order of magnitude, one can 
disregard the $\si$-independent terms in (\ref{beta}).
If we omit also the small term $b_4$, the coefficients of 
the Eq. (\ref{wave}) become constants and we can easily 
check whether the growing modes are present using 
the standard methods. Let us make a 
weaker approximation and regard $\si$ as a big constant.
Then, the Routh-Hurwitz determinants are
$$
D_0 = b_0\,,\,\,\,\,\,\,\,\,\,\,\,\,\,\,\,\,\,\,
D_1 = b_1\,,\,\,\,\,\,\,\,\,\,\,\,\,\,\,\,\,\,\,
D_2 = 60 H^3[a_1+w\cdot\si(t)]^2\,,
$$
which are positive for any parameters of the theory;
$$
D_3 = 360H^6[a_1+w\cdot\si(t)]^3
+432H^6wb[a_1+w\cdot\si(t)]^2\,,
$$
which is indeed positive for $\cdot\si(t)+a_1/w > -6b/5$.
Since $b$ is a negative number of the order (at most) 10,
this condition is satisfied if we do not include too 
large negative coefficient $a_1$ into the classical action 
of vacuum (\ref{higher}). Finally, 
$$
D_4 =  -12bH^4\cdot D_3
$$
is positive if $D_3$ is positive. As we shall see below, 
for the values $n < n_c$, where $n_c$ is a cut-off depending
on $M_*$, this means that the anomaly-induced inflation is 
stable under metric perturbations. This fact is very 
important for the whole model. If we admit that the 
transition from stable to unstable inflation occurs 
at $H = M_* \leq 10^{-5}\,M_P$, then this scale 
will define the maximal amplitude of the gravitational 
waves and our model fits with the standard restrictions
coming from the anisotropy of the CMBR. This conclusion
is achieved without fine-tuning of the parameters of the
model and looks quite robust. In the consideration above we 
have neglected the derivative ${\dot H}$.
This can be justified, for example, using Eq. 
(\ref{replace1}). Remember that, by dimensional 
reason, ${\dot H}$ must be compared to $H^2$. 
A simple calculation gives 
$$
\frac{\dot H}{H^2}
\approx - \frac{{\tilde f}}{2(1-{\tilde f}\si)}\,.
$$
For the SUSY GUT the magnitude of ${\tilde f}$ is at
most $10^{-5}$, while the expression $(1-{\tilde f}\si)$
becomes comparable to $10^{-5}$ only at the very end 
of the stable inflation when the expression (\ref{replace1})
and the whole approach based on the ``sharp-cut-off''
needs to be replaced by a detailed description of 
the decoupling mechanism for massive sparticles in curved 
space-time. Therefore, our approximation
${\dot H}/{H^2}\approx 0$ is reliable in the region 
where the whole approach is reliable, and can be 
considered as a qualitative hint for the 
decoupling region.

The numerical integration of the equation (\ref{wave}) 
have been performed for various magnitudes of the wavenumber
vector $\,n\,$ and the initial conditions (\ref{initi}).
The plots illustrating the stability of $h(t)$ 
are presented at Fig. 4a, where the plots correspond to 
$\,n \leq n_c=0.1$. As we see, there is a very small 
difference between these plots such that the spectrum 
is flat. For the larger values 
of $\,n\,$ the plots have greater growth at the initial 
period (see Fig. 4b) After this initial growth 
the perturbations decrease exponentially according
to our analytical consideration. But, this decrease 
is too slow and for the large $\,n\,$ the waves leave the 
horizon with the amplitude much greater than the 
initial one. The effect is essentially stronger 
for smaller values of the parameter ${\tilde f}$.
The important observation is that, while the exponential 
decrease of the perturbations (after the initial growth)
does not depend on the choice of the vacuum and the 
approximation ${\tilde H}\approx const$, the dangerous 
initial growth may change a lot if we take these aspects 
into account. Therefore, the final conclusion concerning 
the amplitude of the perturbations with large $n$ can be 
achieved only after a more complete analysis. 
The positive result for $\,n<n_c\,$ give a hope to meet an 
acceptable behavior of the metric perturbations.

The main difference with the previous analysis of the 
perturbations \cite{star1,wave} is that here we have to
disregard all the $\si$-independent terms in (\ref{wave}).
This changes the structure of the equation and the 
growing modes disappear. Let us remark that the numerical 
analysis of the metric perturbations in the {\it first} 
$65\,\,e$-folds of inflation shows that, say, the 
perturbations of quantum origin grow up very fast, such 
that the amplitude increases about 15 orders of magnitude. 
Indeed, those huge waves can not be observed and their 
existence does not contradict the CMBR data. However, 
it is very important that the magnitude of the 
perturbations is indeed {\it decreasing} with respect 
to the inflationary background metric. As a result,
the universe enters into the 
{\it last} $65\,\,e$-folds of inflation in a very 
isotropic initial state. All in all, within our 
framework the anomaly-induced inflation does not 
meet unsolvable problems with the gravitational 
waves.

\section{Conclusions and discussions}

$\,\,\,\,\,\,\,$
We have investigated the stability of the anomaly-induced
inflation \cite{star,anju,insusy} with respect to the 
small perturbations.
The stability holds from the initial stage when the 
quantum fields may be approximately considered massless,
until the scale $M_*$, when the most of the sparticles
decouple and the inflation becomes unstable. Thus, this
model of inflation does not need a fine tuning neither 
for the initial data nor for a graceful exit to the FRW
stage. The equations 
for the gravitational waves are essentially different for 
the first and the last $65\,\,e$-folds of the inflation. 
In particular, during the last $65\,\,e$-folds the 
amplitude of the perturbations with $\,n < n_c\,$
does not grow up and the
spectrum of these perturbations is flat. Thus, for the choice 
of the scale $M_*\leq 10^{-5}\,M_P$, the anomaly-induced
inflation may be consistent with the CMBR constraints.

In the period of inflation previous to the transition 
to the unstable stage, we met a rapid oscillations 
of the curvature, very similar to those which were 
discovered earlier by Suen and Anderson for a similar 
(technically) higher-derivative model \cite{andsue}
(see also \cite{suen}). Therefore, 
the anomaly-induced inflation provides an excellent 
possibility to describe the reheating during the last 
stage of the stable inflation and in the transitional 
period. Finally, let us remark that the next important 
steps in the development of this model should be 
the quantitative description of the decoupling 
process and the corresponding cosmological epoch, and 
also derivation and analysis of the density perturbations.
After these two principal problems will be solved, one can 
start to investigate different supersymmetric 
models and their cosmological consequences.

At the present level of understanding the model of 
anomaly-induced inflation looks vert attractive from 
both phenomenological (no fine tuning) and theoretical 
point of view. Let us discuss the last aspect. In the 
modern high energy theoretical physics there are 
several drawbacks. One example is the supersymmetry and 
its low-energy breaking. Even if this breaking is a 
high energy effect such that $M_{SUSY} \gg M_F$, it 
looks unnatural that 
there is not a single couple of the observed particles
which could be superpartners of each other. This 
problem becomes even more explicit for the low-energy
supersymmetry model like MSSM. But, in the 
anomaly-induced inflation we see that exactly this
spectrum of supersymmetry is favored by the scheme
suggested in \cite{insusy}. Moreover, even the 
``less natural'' soft supersymmetry breaking is 
favorable, because for the spontaneous supersymmetry 
breaking
there could be problems with the decoupling of heavy 
sparticles. Another example is the sign universality 
of the vacuum $\be$-functions (\ref{w}) and (\ref{b})
and the variability of sign for (\ref{c}). As we saw 
in sections 4 and 5, the conditions $\,\,b<0\,,\,\,w>0\,\,$ 
are absolutely necessary for the nice stability properties 
of the inflationary model,
while the variability of $\,c\,$ is very helpful for 
the successful realization of the transition between 
stable and unstable regimes \cite{insusy}. It might 
happen that all this indicates to a much more general 
theory which can give us a correct and unified understanding 
of both vacuum quantum effects and the supersymmetry 
breaking. From the effective quantum field theory point 
of view, the anomaly-induced inflation shows that there 
are chances to link these different issues in a new way
and finally approach this general theory.

\vskip 8mm
\noindent
{\large\bf Acknowledgments.}
One of the authors (I.Sh.) is very grateful to J. Fabris,
I.I. Kogan, V.N. Lukash, J. Sola  and A.A. Starobinsky for 
useful discussions, and especially to J. Sola for the 
collaboration on the early stage of the work. He is also 
thankful to his colleagues and friends from the Department 
of Theoretical Physics at the University of Zaragoza (Spain) 
for warm hospitality and stimulating interest to his studies. 
The work of the authors has been supported by the scholarships 
from CNPq, by the research grant from FAPEMIG
and (I.Sh.) by the grant from DGU-MEC (Spain).

\vskip 8mm
\noindent
{\large\bf Appendix.}

This appendix contains a brief summary of 
the stability issues and especially of 
the Routh-Hurwitz conditions, which were extensively 
used in the text of the paper. We refer to the book 
\cite{cezare} for the details. 
Suppose we know a particular solution 
$x_0(t)$ of the equation 
$$
x^{(n)} = f(x^{(n-1)}, x^{(n-2)}, ...,x^\prime, y)
\eqno(A1)
$$
where $x=x(t)$ and $x^{(n)}=\frac{d^nx}{dt^t}$. In order 
to check the asymptotic stability of this solution under 
small perturbations we consider $x(t)=x_0(t)+y(t)$ and 
expand (A1) in the first order in $y(t)$. 
$$
b_0 y^{(n)} + b_1 y^{(n-1)} +
b_2 y^{(n-2)} + ... + b_{n-1} y^\prime + b_n y =0\,,
\eqno(A2)
$$
where $b_0 > 0$ for definiteness. Let us consider two 
possibilities: 

i) All $b_i$ are constants. Then the problem is 
reduced to the algebraic characteristic equation
$$
b_0 \lambda^n + b_1 \lambda^{n-1} +
b_2 \lambda^{n-2} + ... + b_{n-1} \lambda + b_n  = 0\,,
\eqno(A3)
$$
because a general solution of (A2) is a linear
combination of the terms $\,\exp ({\lambda_i\cdot t})$
where $\lambda_i$ are the roots of Eq. (A3) and the 
coefficients are polynomials in $t$. Hence, if all 
these roots have negative real part 
$$
\Re (\lambda_i) < 0\,,
\eqno(A4)
$$
the solution 
$y_0(t)$ is asymptotically stable. A useful criterion of 
(A4) is called the Routh-Hurwitz conditions. The
conditions are that the determinants $D_i$ must be
positive, where $\,D_{1}=b_{1}\,$ and 
$$
 D_k = det \left( 
\begin{array}{ccccc}
b_1  &  b_3  &  b_5  & ... &  b_{2k-1}  \\ 
b_0  &  b_2  &  b_4  & ... &  b_{2k-2}  \\ 
0    &  b_1  &  b_3  & ... &  b_{2k-3}  \\ 
0    &  b_0  &  b_2  & ... &  b_{2k-4}  \\ 
...  & ...   & ...   & ... &  ...       \end{array}
\right)\,,\,\,\,\,\,\,\,\,\,\,\,\,k >1\,,
\eqno(A5)
$$
where $k=2,3,...,n$ and $b_j=0\,,\,\,j>n$. For 
$n=4$ we meet four determinants $D_{1,2,3,4}$.
\vskip 3mm

ii) In the cases $k=1,-1$ of the subsection 3.1 we met the case 
of the non-constant coefficients $b_i=b_i(t)$. But, there is a 
general theorem (see e.g. \cite{cezare}) that the problem of 
asymptotic stability for such case can be reduced to the  
one with the constant coefficients ${\bar b}_i$, provided that 
there are finite limits $\,b_i(t) \to {\bar b}_i\,$ when 
$\,t \to \infty$. As we saw, this is exactly what happens for
$\,k=1,-1$, where $\,{\bar b}_i\,$ correspond to the 
$\,k=0\,$ case.



\newpage

\centerline{\Large \bf Figure captions}
\vskip 5mm

\begin{quotation}
\vskip 1mm
\noindent
{\large\bf Figure 1}.$\,\,$
The plots representing numerical solutions
$\si(t^\prime)$ (where $t^\prime=t/16\pi$) of the equation 
(\ref{central sigma}) for the toy model with ${\tilde f}=10^{-6}$
and MSSM particle content. 
The plot (a) is the same as the one obtained in \cite{shocom}.
In the region ${\dot \si} < M_*$ the two plots (a) and (b) are
equivalent. The plot (b) is almost identical to the parabola
(\ref{parabola}). At the top point of the plot (b) the value
of $\sigma$ equals 1000000, while in case of parabola
$\sigma$ equals 1000001.
\end{quotation}
\vskip 5mm

\begin{quotation}
\vskip 1mm
\noindent
{\large\bf Figure 2}.$\,\,$
The graphics illustrating the behavior of the perturbation 
measure $D$ for the first (a) and last (c) 65 $e$-folds
of inflation. The plot (b) slightly differs from (a),
because of the presence of the matter (dust) with the
density $M_P^4$ at the initial point of inflation.
One can see that the stabilization takes a very short time 
in all cases.
\end{quotation}

\vskip 5mm
\begin{quotation}
\vskip 1mm
\noindent
{\large\bf Figure 3}.$\,\,$
The plot of numerical solution for $H(t)$, in the region 
where the oscillations are explicitly visible.  
\end{quotation}
\vskip 5mm

\vskip 1mm
\begin{quotation}
\noindent
{\large\bf Figure 4}.$\,\,$
The samples of the plots of the numerical solutions $\,h(t)$
in the last $65$ $e$-folds of inflation and initial spectrum 
(\ref{initi}) with the normalization $h(0)=1$
and ${\tilde f}=10^{-5}$.
\end{quotation}

\end{document}